\documentclass[aps,twocolumn,showpacs,groupedaddress]{revtex4}
\usepackage{amssymb,amsmath}
\usepackage{graphicx,array}
\usepackage{dcolumn}
\usepackage{bm}

\begin{document}
\title{Revisiting the zero-temperature phase diagram of stoichiometric SrCoO$_{3}$ with first-principles methods}

\author{Pablo Rivero}
\affiliation{Center for Computation and Technology, Louisiana State University, Baton Rouge, Louisiana 70803, USA}

\author{Claudio Cazorla}
\thanks{Corresponding Author}
\affiliation{School of Materials Science and Engineering, UNSW Australia, Sydney NSW 2052, Australia \\
	     Integrated Materials Design Centre, UNSW Australia, Sydney NSW 2052, Australia}

\begin{abstract}
By using first-principles methods based on density functional theory we revisited the zero-temperature 
phase diagram of stoichiometric SrCoO$_{3}$, a ferromagnetic metallic perovskite that undergoes significant 
structural, electronic, and magnetic changes as its content of oxygen is decreased. We considered both bulk 
and epitaxial thin film geometries. In the bulk case, we found that a tetragonal $P4/mbm$ phase with moderate 
Jahn-Teller distortions and $c/a$ ratio of $\sim 1/\sqrt{2}$ is consistently predicted to have a lower 
energy than the thus far assumed ground-state cubic $Pm\bar{3}m$ phase. In thin films, we found two 
phase transitions occurring at compressive and tensile epitaxial strains. However, in contrast to previous 
theoretical predictions, our results show that: (i)~the phase transition induced by tensile strain is 
isostructural and involves only a change in magnetic spin order (that is, not a metallic to insulator 
transformation), and (ii)~the phase transition induced by compressive strain comprises simultaneous 
structural, electronic and magnetic spin order changes, but the required epitaxial stress is so 
large ($<-6$\%) that is unlikely to be observed in practice. Our findings call for a revision of the 
crystallographic analysis performed in fully oxidised SrCoO$_{3}$ samples at low temperatures, as well 
as of previous first-principles studies. 
\end{abstract}
\pacs{77.55.Nv, 61.50.Ks, 68.65.Cd, 77.80.bn}
\maketitle

\section{Introduction}
\label{sec:intro}
Oxide perovskites with the general formula ABO$_{3}$, in which A and B denote different cations, can 
undergo abrupt structural, magnetic, and electronic changes upon application of small external fields 
(e. g., electric, magnetic, and mechanical) and variation of temperature~[\onlinecite{hatt10,rondinelli11,
rondinelli11b,may10,cazorla14,cazorla15a,wu12,stengel12b,hong13,cazorla13}]. 
This singular reactivity converts oxide perovskites in excellent candidate materials for developing 
new information storage and energy conversion technologies. One illustrative example is the design of 
spintronic devices, in which spin-polarised electric currents are generated with electromagnetic 
fields or through spin-injection from a ferromagnetic material, in order to exploit the magnetic degree of 
freedom of electrons~[\onlinecite{ganichev01,luders06,gajek07}]. In this latter context, ferromagnetic 
metallic perovskites appear to be especially valuable as they can be employed directly as electrodes in 
complex oxide heterostructures.

SrCoO$_{x}$ is a ferromagnetic metallic perovskite in which significant structural, electronic, and 
magnetic order changes occur as its content of oxygen is varied ($2.5 \le x \le 3.0$)~[\onlinecite{jeen13a,jeen13b}]. 
In stoichiometric samples ($x = 3.0$) the crystalline phase that has been repeatedly reported as the ground 
state is the perovskite $Pm\bar{3}m$ structure, which exhibits ferromagnetic (FM) spin order at temperatures 
below $T_{C} = 280-305$~K~[\onlinecite{bezdicka93,kawasaki96,long11}]. Experiments and theoretical calculations 
have assuredly shown that SrCoO$_{3}$ possesses an intermediate spin (IS) state with a magnetic moment of 
$\sim 2.6\mu_{B}$ per Co ion~[\onlinecite{long11,potze95,zhuang98,hoffmann15}]. In non-stoichiometric 
samples with high oxygen deficiency ($x = 2.5$) the system adopts the brownmillerite phase, which presents 
atomically ordered one-dimensional vacancy channels and antiferromagnetic (A) spin order at temperatures 
below $T_{C} = 570$~K~[\onlinecite{takeda72,taguchi79,xie11}].

Recently, it has been shown that the content of oxygen in SrCoO$_{x}$ thin films can be effectively 
tuned by means of epitaxial strain~[\onlinecite{jeen13a,jeen13b,choi13,callori15,hu15,petrie16}]. This 
control of oxygen non-stoichiometry offers a promising new route for designing multifunctional oxide 
materials with improved electrochemical, magnetoresistance, and catalytic properties. Interestingly, Lee 
and Rabe have identified SrCoO$_{3}$ thin films as good candidate materials in which to observe 
large magnetoelectric effects, that is, a strong coupling between polar and magnetic degrees of freedom, 
based on the results of first-principles density functional theory (DFT) calculations~[\onlinecite{lee11,lee11b}]. 
Specifically, it has been predicted that simultaneous magnetic-ferroelectric and metal-insulator transitions 
could be induced in SrCoO$_{3}$ by means of moderate tensile and compressive epitaxial strains. If confirmed,
these results could advance the development of low-power high-efficiency electronic and energy conversion 
technologies.   

Here, we present a revision of the zero-temperature phase diagram of stoichiometric SrCoO$_{3}$ (SCO), 
both in bulk and thin film geometries, using a wide variety of first-principles DFT techniques (i. e., 
LDA+U, GGA+U and hybrid exchange-correlation functionals). Our computational analysis consistently 
shows that, rather than the well-known cubic $Pm\bar{3}m$ phase, a tetragonal $P4/mbm$ phase 
with moderate Jahn-Teller distortions and a $c/a$ ratio of $\sim 1/\sqrt{2}$ is the energetically most 
favorable phase in bulk SCO at zero temperature. We arrived at the same conclusion also when considering 
the application of hydrostatic pressures of up to $70$~GPa. This central outcome either calls for a revision 
of previous crystallographic analysis performed in SCO at low temperatures, or well it represents a failure 
of DFT methods in describing the competition among crystalline phases in this highly correlated material. 
We have also investigated the energy, magnetic, and electronic properties of SCO thin films in a wide interval 
of epitaxial strains, $-7$\%~$\le \eta \le +5$\% ($\eta \equiv \left(a-a_{0}\right)/a_{0}$, where $a_{0}$ is 
the in-plane equilibrium distance between cations of the same species), using computational techniques similar 
to those employed by Lee and Rabe (i. e., DFT GGA+U)~[\onlinecite{lee11,lee11b}]. Our results show a phase-transition 
scenario that is compatible with recent experimental observations but appreciably different from the one proposed 
previously. Actually, our findings lower the expectations of realising large magnetoelectricity in stoichiometric 
SCO thin films.

The organisation of this article is as it follows. In the following section we provide an overview 
of the computational methods employed in this study. Next, we present our results along with 
some discussion. Finally, we summarise our main findings in Sec.~\ref{sec:summary}.

\begin{figure}
\centerline{
\includegraphics[width=1.0\linewidth]{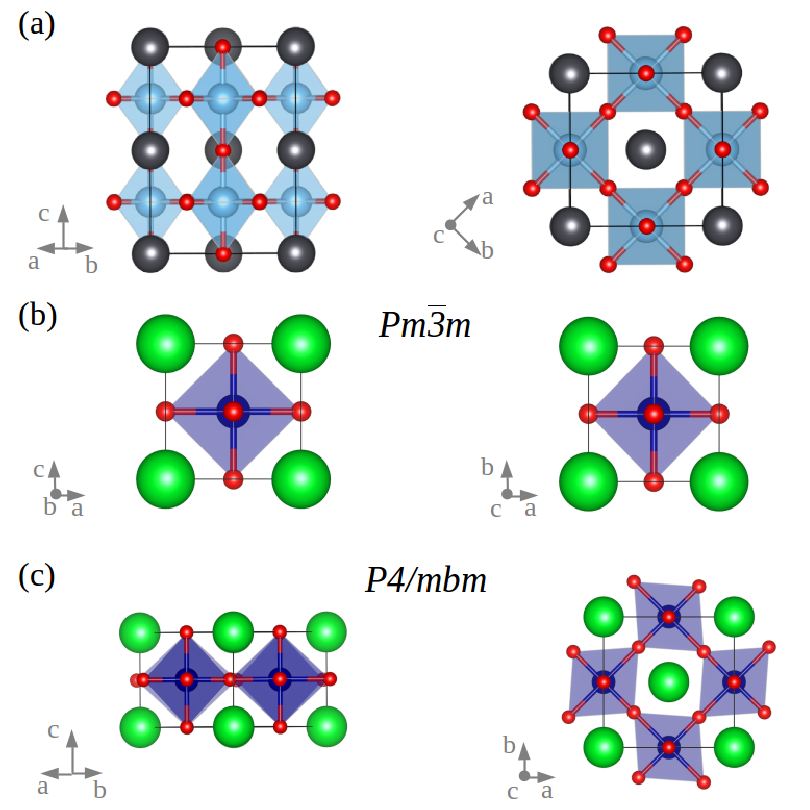}}
\caption{(a)~Sketch of the $20$-atom $\sqrt{2} \times \sqrt{2} \times 2$ simulation cell  
         used in most of our calculations; black, blue, and red spheres represent A, B, and 
         O atoms in ABO$_{3}$ perovskites. The corresponding lattice vectors are ${\bf a_{1}} 
         = (a, a, 0)$, ${\bf a_{2}} = (a, -a, 0)$, and ${\bf a_{3}} = (0, 0, 2a)$, where $a$ 
         is the lattice parameter of the $5$-atom primitive perovskite cell. (b)~Sketch of 
         the cubic $Pm\bar{3}m$ phase. (c)~Sketch of the tetragonal $P4/mbm$ phase. Sr, Co, 
         and O atoms are represented with green, blue, and red spheres, respectively.}
\label{fig1}
\end{figure}

\section{Computational Methods}
\label{sec:methods}
In order to accurately describe the physical properties of SrCoO$_{3}$ it is necessary to use 
methods that go beyond standard DFT, e. g., GGA+U (or LDA+U) and hybrid functionals, as due to the 
presence of strongly correlated $d$ electrons. The GGA+U approximation is based on a Hubbard-type 
Hamiltonian approach in which an on-site constant effective potential is introduced to account 
for the intra-atomic electron Coulomb repulsive ($J_{1}$) and intra-orbital exchange potential 
($J_{2}$)~[\onlinecite{liechtenstein95}]. Here we employ the GGA+U (and LDA+U) scheme due to 
Dudarev~[\onlinecite{dudarev98}] for a better treatment of Co's $3d$ electrons; the meaningful 
term in this approach is U$_{eff} \equiv J_{1} - J_{2}$, to which we refer hereafter simply as 
to U. The choice of the U value, however, is not unique or rigorously well defined. In general, 
this is selected so as to reproduce a particular set of experimental data (e. g., lattice 
parameters or energy band gaps) or theoretical results obtained with higher accuracy methods 
as closely as possible~[\onlinecite{hernandez09,hong12,rivero16}]. 

Hybrid functionals represent a good alternative to DFT+U approximations. In this context, one 
mixes a particular amount of nonlocal Hartree-Fock (HF) exchange, $\alpha$, with standard density 
functional (either LDA or GGA) exchange potentials. The resulting hybrid DFT exchange-correlation 
potential can be expressed as:
\begin{equation}
V_{xc}^{hybrid} = \alpha V_{x}^{HF} + (1 - \alpha) V_{x}^{GGA/LDA} + V_{c}^{GGA/LDA}~,  
\label{eq:hybrid}
\end{equation} 
where $V_{x}$ and $V_{c}$ represent exchange and correlation functionals, respectively. In contrast
to GGA+U and LDA+U, hydrid functionals account for both the electronic localisation acting on all
the states of the system and nonlocal exchange effects. Yet, in a strict \emph{ab initio} sense, 
there is not a general prescription for chosing the right amount of nonlocal HF exchange. 
   
In this work, we have used both the GGA+U (LDA+U) and hybrid functional approaches. Next, we explain 
the technical details of our calculations. 

\subsection{GGA+U calculations}
\label{subsec:gga+u}
In most of our calculations we used the generalised gradient approximation to density functional theory (DFT) proposed 
by Perdew, Burke, and Ernzerhof (PBE)~[\onlinecite{pbe96}], as it is implemented in the VASP package~[\onlinecite{vasp}]. 
Other common exchange-correlation functional approximations such as LDA~[\onlinecite{ceperley80}] and PBE$^{\rm sol}$~[\onlinecite{pbesol}] 
were also considered. We used the ``projector augmented wave'' method to represent the ionic cores~[\onlinecite{bloch94}], 
considering the following electrons as valence states: Co's $3p$, $3d$, and $4s$; Sr's $4s$, $5s$, and $4p$; and O's $2s$ 
and $2p$. The wave functions were expanded in a plane-wave basis truncated at $500$~eV, and a $\sqrt{2} \times \sqrt{2} 
\times 2$ simulation cell containing up to $20$ atoms was used in most of our energy and geometry relaxation calculations 
(see Fig.~\ref{fig1}a). For integrations in the Brillouin zone (BZ), we employed $\Gamma$-centered ${\rm q}$-point grids 
of $8 \times 8 \times 8$. Using these parameters we obtained enthalpy energies that were converged to within $0.5$~meV per 
formula unit (f.u.). Geometry relaxations were performed using a conjugate-gradient algorithm that varied the shape and in 
some cases also the volume of the unit cell; the imposed tolerance on the atomic forces was $0.01$~eV$\cdot$\AA$^{-1}$. 
High-pressure equations of state were determined by computing the total energy of the crystal in a series of volume points 
that subsequently were fitted to analytical Birch-Murnaghan functions~[\onlinecite{cazorla15b}].

We also calculated the vibrational phonon spectrum of several bulk phases by using the ``direct method''~[\onlinecite{kresse95,alfe01}]
and DFT calculations. In the direct method, the force-constant matrix of the crystal is calculated in real-space
by considering the proportionality between atomic displacements and forces when the former are sufficiently small.
Large supercells need to be constructed in order to guarantee that the elements of the force-constant matrix have
all fallen off to negligible values at their boundaries, a condition that follows from the use of periodic boundary
conditions~[\onlinecite{alfe09}]. Once the force-constant matrix is calculated one can Fourier-transform it to
obtain the phonon spectrum at any ${\bf q}$-point. The quantities with respect to which our phonon calculations need 
to be converged are the size of the supercell, the size of the atomic displacements, and the numerical accuracy in 
the sampling of the Brillouin zone. We found the following settings to provide zero-point energies~[\onlinecite{cazorla13}] 
converged to within $5$~meV/f.u.: $2 \times 2 \times 2$ supercells containing up to $160$ atoms, atomic displacements of 
$0.02$~\AA, and ${\rm q}$-point grids of $12 \times 12 \times 12$. The value of the phonon frequencies were obtained with 
the PHON code developed by Alf\`e~[\onlinecite{alfe09}]. In using this code we exploited the translational invariance
of the system to impose the three acoustic branches to be exactly zero at the $\Gamma$ ${\bf q}$-point, and used central
differences in the atomic forces (i. e., positive and negative atomic displacements were considered).

\subsection{Hybrid functional calculations}
\label{subsec:hybrid}
Hybrid functional DFT calculations were performed with the replicated-data version of the CRYSTAL14 package~[\onlinecite{crystal14}]. 
This is a first-principles electronic structure software which employs atom-centered Gaussian-type orbital (GTO) basis 
sets to build Bloch functions that represent the one-electron crystalline orbitals. GTO offer a number of convenient 
computational features, e. g., the use of local basis sets containing minimal overlap with neighboring orbitals, that 
allow to perform HF exchange calculations in medium and large size systems affordably (e. g., the resulting computational 
expense scales as $N^{2-3}$ with the number of particles, to be compared with the usual $N^{4}$ scaling found in plane-waves 
based methods). All-electron GTO atomic basis sets were chosen as it follows: for Co's we used the double-zeta 
all-electron basis set from~[\onlinecite{webpage}]; for Sr's small-core Hay-Wadt pseudopotentials~[\onlinecite{hay85}] 
were adopted for the description of the inner-shell electrons $1s$, $2s$, $2p$, $3s$, $3p$, and $3d$, while for the valence 
part $4s$, $4p$, and $5s$ we used the optimized basis set successfully applied to the strontium titanate 
study~[\onlinecite{piskunov04}]; and for O's we used the $8-411d$ all-electron basis set constructed by Cor\`a~[\onlinecite{cora05}]. 

A Monkhorst-Pack $8 \times 8 \times 8$ ${\rm q}$-point grid was used for BZ sampling. The thresholds controlling the accuracy 
in the calculation of the Coulomb and exchange integrals were set equal to 10$^{-7}$ and 10$^{-14}$, and to 10$^{-7}$~eV in 
the SCF energy. The crystal cell parameters and atomic positions were relaxed during the geometry optimisations by imposing a 
convergence criterion of $0.008$~eV$\cdot$\AA$^{-1}$ in the atomic forces. The hybrid functionals employed in this study 
include PBE-10~[\onlinecite{rivero16}], PBE0~[\onlinecite{pbe0}], and HSE06~[\onlinecite{hse06}]. In the first two versions, 
the amount of HF exchange is $0.10$ and $0.25$, respectively (that is, parameter $\alpha$ in Eq.~\ref{eq:hybrid}); in the HSE06 
case, there is a separation between short and long ranges in which $\alpha = 0.25$ and $1.00$ are used, respectively.

\section{Results and Discussion}
\label{sec:results}

\begin{figure}
\centerline
        {\includegraphics[width=0.90\linewidth]{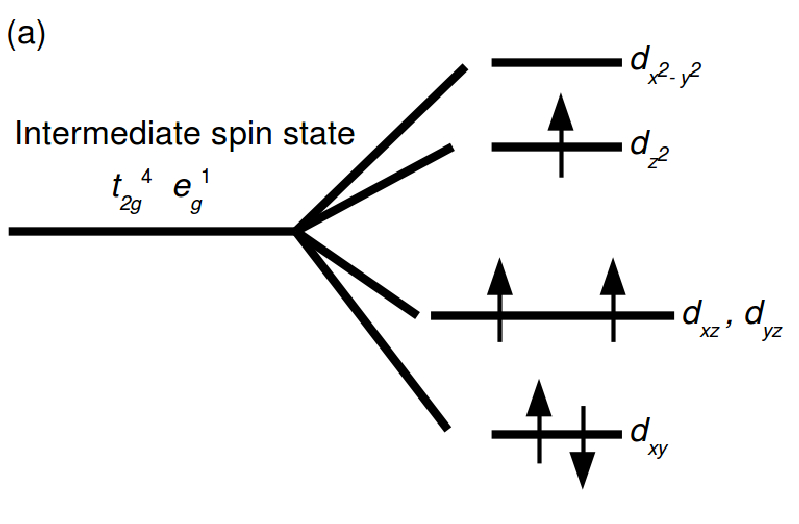}}
        {\includegraphics[width=1.00\linewidth]{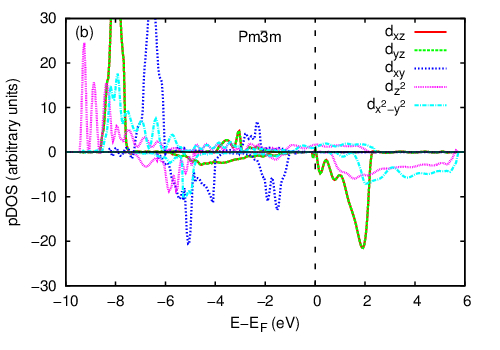}}
\caption{(a)~Sketch of a possible intermediate spin state expected to occur in bulk SrCoO$_{3}$~[\onlinecite{potze95}].
         (b)~Density of electronic $d$ states calculated for bulk SrCoO$_{3}$ in a cubic $Pm\bar{3}m$ phase with the 
         PBE-10 hybrid functional. Spin-up and spin-down electronic densities are represented with positive and negative 
         values, respectively.}
\label{fig2}
\end{figure}

\subsection{Intermediate spin state and the choice of U}
\label{subsec:ismodel}
There is strong experimental and theoretical evidence showing that bulk SCO possesses an intermediate spin (IS) state resulting 
from a competition between intra-atomic exchange interactions and the crystal field~[\onlinecite{long11,potze95,zhuang98,hoffmann15}]. 
This IS configuration can be understood as a high spin state in the Co$^{3+}$ ions that is antiferromagnetically coupled to 
a ligand hole of $e_{g}$ symmetry, which can be formally represented with the $d$-orbital occupation model $t^{4}_{2g}e^{1}_{g}$ 
(see Fig.~\ref{fig2}a)~[\onlinecite{potze95,hoffmann15}]. In Fig.~\ref{fig2}b, we show the partial density of electronic 
$d$ states (pDOS) calculated in bulk SCO in the cubic $Pm\bar{3}m$ phase with the hybrid functional PBE-10 (see Fig.~\ref{fig1}b); 
our hybrid functional pDOS results reproduce the expected orbital occupation configuration $t^{4}_{2g}e^{1}_{g}$: two unoccupied 
$t_{2g}^{\downarrow}$ states ($d_{xz}$ and $d_{yz}$ above the Fermi energy level, $E_{F}$) and one occupied $t_{2g}^{\downarrow}$ 
state ($d_{xy}$ below $E_{F}$), while $e_{g}$ states are smeared over a large energy range containing $E_{F}$ in the spin-up 
channel. 

\begin{figure}
\centerline{
\includegraphics[width=1.0\linewidth]{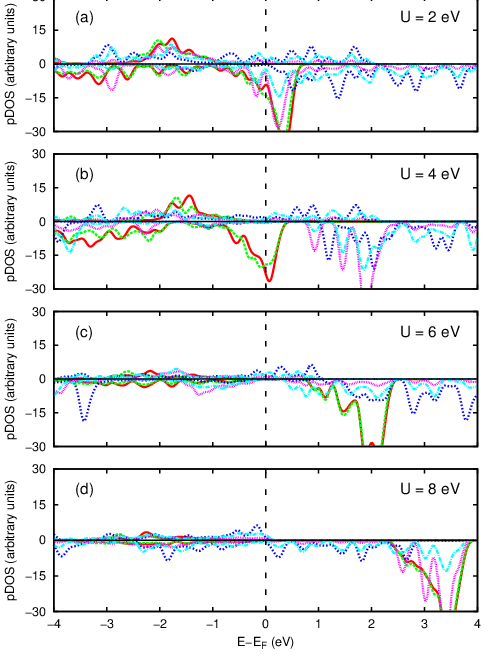}}
\caption{Density of electronic $d$ states calculated in bulk SrCoO$_{3}$ in the cubic $Pm\bar{3}m$ phase
         with the PBE+U method expressed as a function of U. Spin-up and spin-down electronic densities are 
         represented with positive and negative values, respectively. The convention in representing the $d$ 
         states is the same than in the previous figure. U values smaller than $5$~eV do not reproduce the 
         expected intermediate spin state in bulk SCO.}
\label{fig3}
\end{figure}

\begin{table*}
\begin{center}
\label{tab:data}
\begin{tabular}{c c c c c c c c c}
\hline
\hline
$ $ & $ $ & $ $ & $ $ & $ $ & $ $ & $ $ & $ $ & $ $ \\
$ $ & $\quad \Delta E~({\rm meV/f.u.}) \quad$ & $\quad M~(\mu_{B}) \quad$ & $\quad a$~(\AA) \quad & $\quad b$~(\AA) \quad & $\quad c$~(\AA) \quad & $\quad x \quad$ & $\quad Q_{2}$~(\AA) \quad & $\quad Q_{3}$~(\AA) \quad \\
$ $ & $ $ & $ $ & $ $ & $ $ & $ $ & $ $ & $ $ & $ $ \\
\hline
$ $ & $ $ & $ $ & $ $ & $ $ & $ $ & $ $ & $ $ & $ $ \\
${\rm HSE06}  $      & $-632$ & $2.53$ & $5.429$ & $5.429$ & $3.912$ & $0.272$ & $0.474$ & $0.060$ \\
${\rm PBE0}   $      & $-269$ & $2.50$ & $5.424$ & $5.424$ & $3.909$ & $0.227$ & $0.498$ & $0.060$ \\
${\rm PBE-10} $      & $-259$ & $2.28$ & $5.392$ & $5.392$ & $3.979$ & $0.233$ & $0.376$ & $0.136$ \\
$ $ & $ $ & $ $ & $ $ & $ $ & $ $ & $ $ & $ $ & $ $ \\
\hline
$ $ & $ $ & $ $ & $ $ & $ $ & $ $ & $ $ & $ $ & $ $ \\
${\rm PBE+U}       $ & $-126$ & $2.84$ & $5.502$ & $5.502$ & $3.943$ & $0.268$ & $0.397$ & $0.042$ \\
${\rm PBE^{sol}+U} $ & $-117$ & $2.78$ & $5.404$ & $5.404$ & $3.873$ & $0.266$ & $0.345$ & $0.042$ \\
${\rm LDA+U}       $ & $-109$ & $2.86$ & $5.313$ & $5.313$ & $3.811$ & $0.265$ & $0.324$ & $0.044$ \\
$ $ & $ $ & $ $ & $ $ & $ $ & $ $ & $ $ & $ $ & $ $ \\
\hline
\hline
\end{tabular}
\end{center}
\caption{Energy, structural, and magnetic properties of bulk equilibrium SrCoO$_{3}$ calculated in the tetragonal 
        $P4/mbm$ phase using several hybrid and DFT functionals (${\rm U} = 6$~eV). $\Delta E \equiv E(P4/mbm) 
        - E(Pm\bar{3}m)$, $M$ is the magnetic moment per Co ion, $a$, $b$, $c$, the length of the lattice vectors, 
        $x$ the displacement of the equatorial O atoms sitting in $h$ Wyckoff positions, and $Q_{2}$ and $Q_{3}$ 
        lattice Jahn-Teller distortion parameters (see text).}
\end{table*}

In Fig.~\ref{fig3}, we show the $d$-orbital pDOS calculated also in bulk SCO considering the cubic $Pm\bar{3}m$ phase but 
employing the GGA+U method. Several U values were employed in order to analyse the effect of this term on the description 
of the expected IS state. It is found that only for ${\rm U} > 5$~eV the IS state in bulk SCO is reproduced correctly, 
as otherwise a low spin state is obtained (that is, no particular $d$ state is unoccuppied). For instance, in the ${\rm U} 
= 6$~eV case the two unoccupied $t_{2g}^{\downarrow}$ states ($d_{xz}$ and $d_{yz}$ above $E_{F}$) are properly rendered 
whereas in the ${\rm U} = 2$ and $4$~eV cases those appear as partially occupied. We note that these conclusions are in  
agreement with recent results reported by Hoffmann \emph{et al.} in work [\onlinecite{hoffmann15}]. In view of these 
findings, we adopted ${\rm U} = 6$~eV for the rest of our GGA+U (and LDA+U) calculations. It is worth noticing that in 
previous GGA+U studies by Lee and Rabe~[\onlinecite{lee11,lee11b}] ${\rm U} < 5$~eV values were employed, hence the 
$d$-orbital configurations sampled therein are likely to differ from the expected IS state.

\subsection{Bulk SrCoO$_{3}$ at equilibrium}
\label{subsec:bulk}

\begin{figure}
\centerline{
\includegraphics[width=1.0\linewidth]{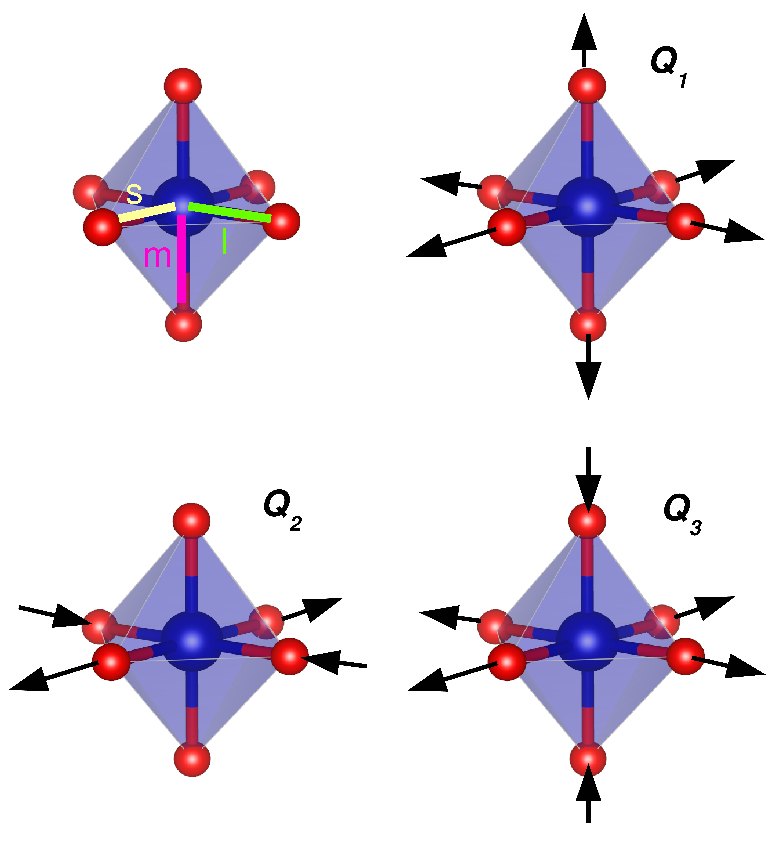}}
\caption{Sketch of possible Jahn-Teller distortion modes in SrCoO$_{3}$ and 
         of the distorted Co-O bonds $l$, $m$, and $s$.}
\label{fig4}
\end{figure}

\begin{figure}
\centerline
        {\includegraphics[width=0.90\linewidth]{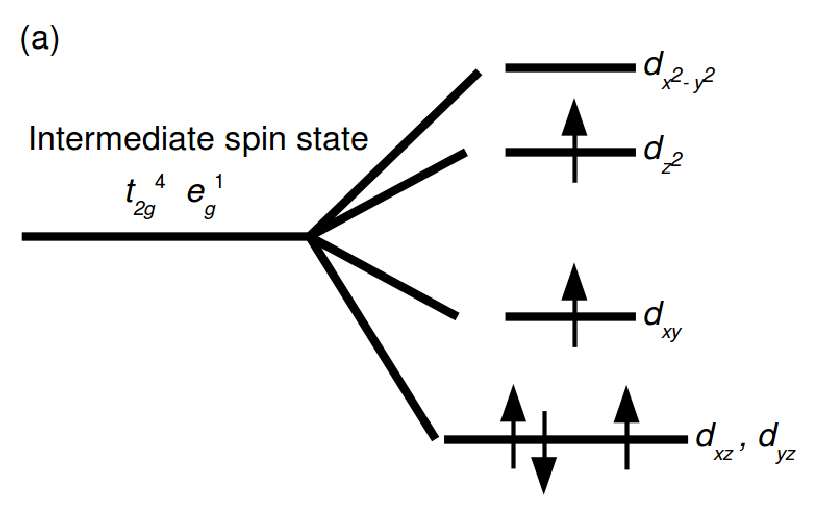}}
        {\includegraphics[width=1.00\linewidth]{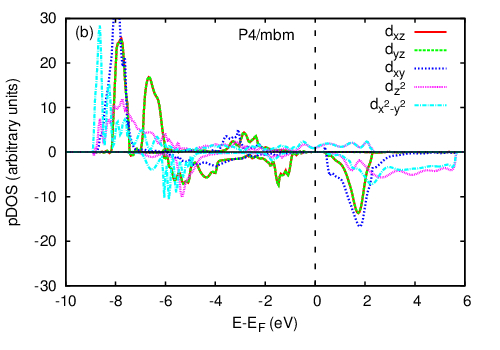}}
\caption{(a)~Sketch of a possible intermediate spin state expected to occur in bulk SrCoO$_{3}$~[\onlinecite{potze95}].
         (b)~Density of electronic $d$ states calculated for bulk SrCoO$_{3}$ in a tetragonal $P4/mbm$ phase with the 
         PBE-10 hybrid functional. Spin-up and spin-down electronic densities are represented with positive and negative 
         values, respectively.}
\label{fig5}
\end{figure}

\begin{figure}
\centerline{
\includegraphics[width=1.0\linewidth]{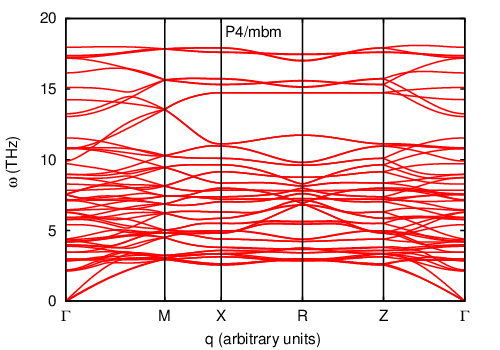}}
\caption{Lattice phonon spectrum calculated for bulk SrCoO$_{3}$ in a tetragonal $P4/mbm$ phase with the PBE+U method
         (${\rm U} = 6$~eV).}
\label{fig6}
\end{figure}

In all our GGA+U, LDA+U and hybrid functional DFT calculations, we consistently found that a tetragonal $P4/mbm$ phase 
displaying moderate Jahn-Teller distortions and a $c/a$ ratio of $\sim 1/\sqrt{2}$ (see Fig.~\ref{fig1}c) has a lower 
energy than the cubic $Pm\bar{3}m$ phase, thus far assumed to be the ground-state in bulk SCO~[\onlinecite{bezdicka93,
kawasaki96,long11}]. In particular, the new tetragonal $P4/mbm$ phase (space group $127$) has a $10$-atoms unit cell in 
which two Co atoms site on $c$ Wyckoff positions, two Sr atoms on $a$ Wyckoff positions, two apical O atoms on $d$ 
Wyckoff positions, and four equatorial O atoms on $h$ Wyckoff positions. The Jahn-Teller (JT) distortions that are 
relevant to this structure can be expressed as: 
\begin{eqnarray}
Q_{2} = \frac{2 \left( l - s \right)}{\sqrt{2}}~, \\
Q_{3} = \frac{2 \left( 2m - l - s \right)}{\sqrt{6}}~, 
\label{jtmodes}
\end{eqnarray}
where $l$, $m$, and $s$ refer to the long, medium and short Co-O distances, respectively (see Fig.~\ref{fig4}). A third
possible JT distortion associated to a breathing mode, $Q_{1}$ (see Fig.~\ref{fig4}), was found to be negligible in 
bulk SCO. 

In Table~I, we summarise our energy, structural, and magnetic results obtained in the new tetragonal $P4/mbm$ phase. 
As it is appreciated therein, the energy difference between the tetragonal $P4/mbm$ and cubic $Pm\bar{3}m$ phases, 
$\Delta E \equiv E(P4/mbm) - E(Pm\bar{3}m)$, is of the order of $\sim 0.1$~eV/f.u., that is, fairly large, with hybrid 
functionals providing the largest $|\Delta E|$ values. In most cases the predicted magnetic moment per Co ion is of 
$2.5-2.9\mu_{B}$, which is consistent with the expected intermediate spin state~[\onlinecite{long11,potze95,zhuang98,hoffmann15}]. 
Regarding the structural features, all DFT functionals render a $c/a$ ratio of $\sim 1/\sqrt{2}$, which denotes very 
small tetragonality as the unit cell of the $P4/mbm$ phase can be thought of a $\sqrt{2} \times \sqrt{2} \times 1$ 
supercell constructed with the usual $5$-atom perovskite unit cell. The $x$ displacement associated to $h$ Wyckoff 
positions, which are occupied by equatorial oxygen atoms, varies from $0.23$ to $0.27$ depending on the employed  
functional. Finally, the $Q_{2}$ JT distortion, which mostly affects to the equatorial plane in the oxygen octahedra 
(see Fig.~\ref{fig4}), appears to be the dominant one.     

We have calculated the $d$-orbital pDOS of bulk SCO in the new tetragonal $P4/mbm$ phase with the hybrid functional 
PBE-10 (see Fig.~\ref{fig5}b). Our results reproduce the expected orbital occupation configuration $t^{4}_{2g}e^{1}_{g}$: 
two unoccupied $t_{2g}^{\downarrow}$ states ($d_{xy}$ and $d_{yz}/d_{xz}$ above $E_{F}$) and one occupied $t_{2g}^{\downarrow}$ 
state ($d_{yz}/d_{xz}$ below $E_{F}$), while $e_{g}$ states are smeared over a large energy range containing $E_{F}$ in the 
spin-up channel. The main effect deriving from the presence of JT distortions and lowering of crystal symmetry as compared 
to the cubic $Pm\bar{3}m$ phase, is to swap the energy ordering between $t_{2g}$ $d$-states (see Figs.~\ref{fig2}a 
and~\ref{fig5}a). Analogous $d$-orbital pDOS results were obtained also whith the PBE+U method.  

We calculated the lattice phonon spectrum of bulk SCO in the new tetragonal $P4/mbm$ phase with the PBE+U method. It 
was found that this structure is vibrationally and mechanically stable as none imaginary phonon frequency appeared in the 
corresponding Brillouin zone (see Fig.~\ref{fig6}). Interestingly, we performed analogous phonon calculations in the cubic 
$Pm\bar{3}m$ phase and found few imaginary phonon frequencies at the high-symmetry reciprocal-space point ${\rm M} = (\frac{1}{2} 
\frac{1}{2} 0)$. This finding is consistent with the fact that by enlarging the $5$-atom perovskite unit cell along the 
$x-y$ Cartesian directions it is possible to find a lower-energy crystal structure at zero temperature. Also, it suggests 
that SCO in the cubic $Pm\bar{3}m$ phase may be highly anharmonic. Consequently, at $T \neq 0$~K conditions, the cubic 
$Pm\bar{3}m$ phase could be entropically stabilised over the tetragonal $P4/mbm$ phase~[\onlinecite{huang01,cazorla12}]; 
such a possibility, however, has not been explored in this work.  

\begin{figure}
\centerline
        {\includegraphics[width=1.00\linewidth]{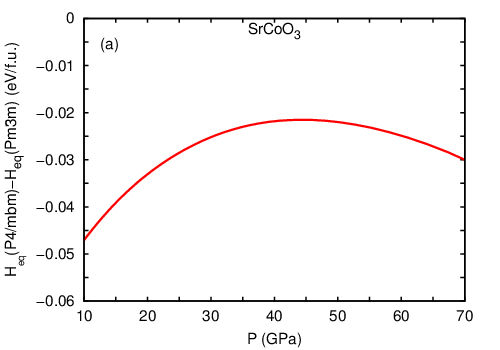}}
        {\includegraphics[width=1.00\linewidth]{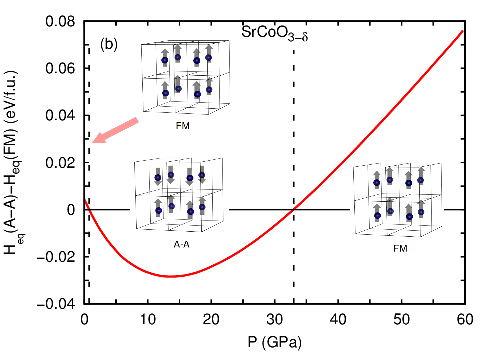}}
\caption{(a)~Enthalpy energy difference between the tetragonal $P4/mbm$ and cubic $Pm\bar{3}m$ phases
          calculated in stoichiometric SCO as a function of pressure.
         (b)~Enthalpy energy difference between magnetic spin orders A-A and FM calculated in non-stoichiometric
         SCO (SrCoO$_{2.75}$) as a function of pressure. The vertical dashed lines indicate $P$-induced
         phase transitions affecting the magnetic spin order.}
\label{fig7}
\end{figure}

\subsection{Bulk SrCoO$_{3}$ under pressure}
\label{subsec:underP}
Under compression, oxide perovskites may undergo phase transformations into higher symmetry 
configurations~[\onlinecite{andrault91,iniguez02}]. For this reason, we analysed the possibility 
of a $P$-induced $P4/mbm ~\to~ Pm\bar{3}m$ phase transition in bulk SCO with the PBE+U 
method. In Fig.~\ref{fig7}a, we show the zero-temperature enthalpy difference calculated 
between the two phases, $\Delta H \equiv H(P4/mbm) - H(Pm\bar{3}m)$, expressed as a function 
of pressure. As it is observed therein, the effect of compression initially is to reduce the 
value of $|\Delta H|$, however, at $P > 45$~GPa this trend is reverted. Such a change of tendency 
is due to the disappearance of an energy band gap in the spin-down channel of the cubic $Pm\bar{3}m$ 
phase (not shown here); the $d$-orbital pDOS estimated in the tetragonal $P4/mbm$ phase, on the 
contrary, always displays an intermediate spin state in the interval $0 \le P \le 70$~GPa.   

In a recent experimental study on bulk SCO at $T = 200$~K, Yang \emph{et al.} have reported two 
$P$-induced phase transitions occurring at $P \sim 1$ and $\sim 45$~GPa and involving a spin reorientation 
and spin-state change, respectively~[\onlinecite{yang15}]. These observations are not coincident
with our results obtained in bulk stoichiometric SCO. The reasons behind such a disagreement 
could be explained in terms of (i)~the neglection of thermal excitations in our theoretical study, 
and/or (ii)~the presence of small oxygen deficiencies in the experimental samples (i. e., 
SrCoO$_{3-\delta}$ with $\delta = 0.05$, as reported by Yang \emph{et al.}~[\onlinecite{yang15}]). 
In order to substantiate the effects of small non-stoichiometry in bulk SCO under pressure (at zero
temperature), we calculated the enthalpy of SrCoO$_{2.75}$ crystals (that is, generated by removing 
one oxygen atom from the stoichiometric $20$-atoms unit cell) as a function of pressure. We 
considered four possible magnetic spin arrangements: ferromagnetic (FM), antiferromagnetic A-type 
(A-A, parallel in-plane spins, antiparallel out-of-plane spins), antiferromagnetic G-type (A-G, 
antiparallel in-plane spins, antiparallel out-of-plane spins), and antiferromagnetic C-type 
(A-C, antiparallel in-plane spins, parallel out-of-plane spins).    

In Fig.~\ref{fig7}b, we plot the zero-temperature enthalpy difference calculated among the 
lowest-energy SrCoO$_{2.75}$ configurations obtained when constraining ferromagnetic (FM) 
and antiferromagnetic A-type (A-A) spin orders, $\Delta H^{m} \equiv H^{A-A} - H^{FM}$, as those 
were found to be the energetically most favorable cases. It is observed that at $P \sim 0$~GPa 
the enthalpy difference $\Delta H^{m}$ adopts very small and positive values, which indicates 
that FM spin order remains the most stable; however, when pressure is increased beyond $0.75$~GPa 
the system with A-A spin order becomes the ground state. We note that this transition point is 
very close to the pressure at which Yang \emph{et al.} have observed a reorientation of magnetic 
spins in SCO (i. e., $1.1$~GPa~[\onlinecite{yang15}]); nevertheless, our calculated $\Delta H^{m}$ 
values within the pressure interval $0 \le P \le 1$~GPa are so small (i. e., of the order of our accuracy 
threshold of $\sim 1$~meV/f.u.) that we cannot discard that bulk SrCoO$_{2.75}$ is already 
antiferromagnetic close to equilibrium. As pressure is increased beyond $33$~GPa, the system 
displaying FM spin order clearly becomes the one with the lowest energy. According to our calculations, 
this magnetic spin order phase transformation is accompanied by a very small volume reduction 
of $0.8$~\%, in which the volume of the A-A phase is $49.15$~\AA$^{3}$.  

In view of our zero-temperature results obtained in bulk SrCoO$_{3-\delta}$ under pressure, Yang 
\emph{et al.}'s observations reported at $P \sim 45$~GPa~[\onlinecite{yang15}] could be reinterpreted 
as a full magnetic spin order transformation, rather than as a spin-state change. In fact, the phase 
transition that we report in SrCoO$_{2.75}$ is isostructural and continuous, as it has been observed 
in the laboratory. However, no intermediate to low spin state transition is observed in our calculations, 
neither in the stoichiometric nor in the nonstoichiometric case, when considering the lowest-energy phases. 
Further experimental studies in compressed SCO are highly desirable in order to rigorously confirm or 
reject our hypothesis.  

\begin{figure*}
\centerline{
\includegraphics[width=0.9\linewidth]{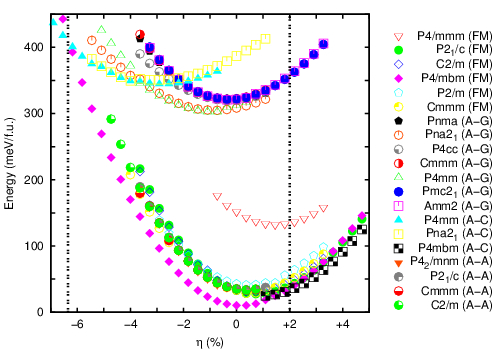}}
\caption{Energy of several competitive phases in SrCoO$_{3}$ thin films calculated as a function 
         of epitaxial strain. The vertical dashed lines indicate $\eta$-induced phase transitions
         affecting the structural, magnetic, and electronic properties of the system.}
\label{fig8}
\end{figure*}

\subsection{SrCoO$_{3}$ thin films}
\label{subsec:thinfilms}

\begin{figure}
\centerline
        {\includegraphics[width=1.00\linewidth]{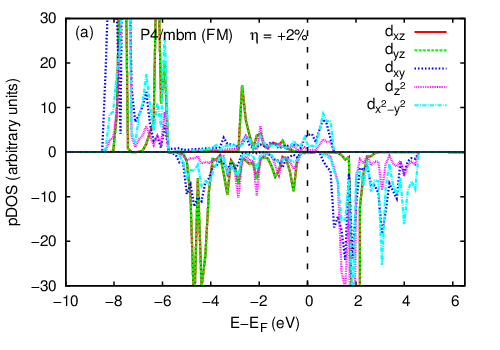}}
        {\includegraphics[width=1.00\linewidth]{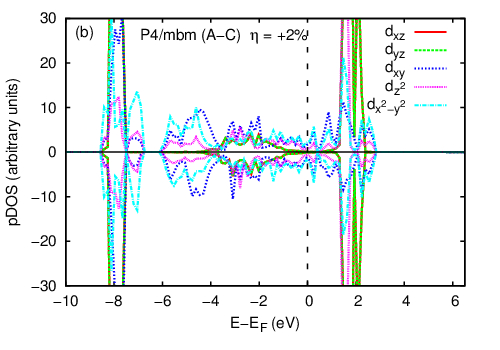}}
\caption{Density of electronic $d$ states calculated with the PBE+U method (${\rm U} = 6$~eV) in 
         tensile SrCoO$_{3}$ thin films in the tetragonal (a)~$P4/mbm$ [FM] and (b)~$P4/mbm$ [A-C] 
         phases. Spin-up and spin-down electronic densities are represented with positive and 
         negative values, respectively.}
\label{fig9}
\end{figure}

\begin{figure}
\centerline
        {\includegraphics[width=1.00\linewidth]{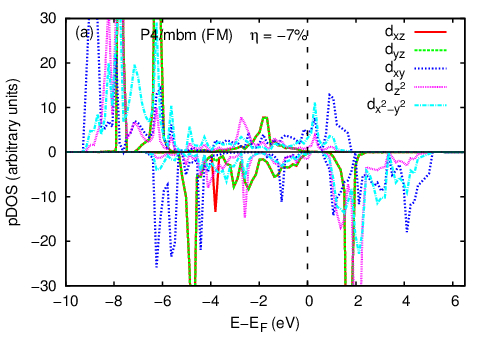}}
        {\includegraphics[width=1.00\linewidth]{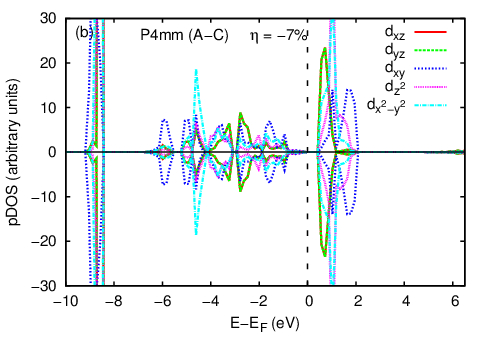}}
\caption{Density of electronic $d$ states calculated with the PBE+U method (${\rm U} = 6$~eV) in 
         highly compressive SrCoO$_{3}$ thin films in the tetragonal (a)~$P4/mbm$ [FM] and (b)~$P4mm$ 
         [A-C] phases. Spin-up and spin-down electronic densities are represented with positive and 
         negative values, respectively.}
\label{fig10}
\end{figure}

Lee and Rabe have recently predicted a series of intriguing multiferroic phase transformation occurring 
in stoichiometric SCO thin films, based on GGA+U DFT calculations~[\onlinecite{lee11,lee11b}]. In particular, 
it has been proposed that both small tensile and compressive epitaxial strains (i. e., of $\sim +2$~\% and 
$\sim -1$~\%, respectively) can trigger a transformation from the bulk FM-metallic phase into an 
A-insulating-ferroelectric phase. These theoretical results suggest that large magnetoelectric effects, that 
is, cross responses to applied electric and magnetic fields, could be realised in SCO thin films, as in 
regions where several multiferroic phases are energetically competitive those can be expected to 
happen~[\onlinecite{jorge08,jorge10}]. In two recent experimental studies on SrCoO$_{3-\delta}$ thin films 
by Callori \emph{et al.}~[\onlinecite{callori15}] and Hu \emph{et al.}~[\onlinecite{hu15}], the existence of 
an antiferromagnetic phase at moderate tensile strains of $\sim +2.7$~\%~ has been ascertained; however, the 
accompanying metal-to-insulator and nonpolar-to-ferroelectric phase transitions as predicted by Lee and Rabe 
appear to be missing. The second multiferroic phase transformation that has been anticipated to occur at 
compressive strains also remains experimentally unverified, in spite of the small epitaxial distortions 
that have been suggested to be involved (i. e., $\sim -1$~\%)~[\onlinecite{lee11,lee11b}].      

In view of the results presented in previous sections, we considered as opportune to revise the zero-temperature 
phase diagram of stoichiometric SCO thin films that is obtained with first-principles methods. We employed the 
PBE+U method with ${\rm U} = 6$~eV, in order to reproduce the expected intermediate spin state correctly (see 
Sec.~\ref{subsec:ismodel}). In addition to the tetragonal $P4/mbm$ and cubic $Pm\bar{3}m$ phases considered so 
far in this study (we note that the latter phase transforms into tetragonal $P4/mmm$ when applying epitaxial 
strain on it), we analysed many other structures exhibiting tetragonal (e. g., $P4mm$, $P4cc$, and $P4_{2}/mnm$), 
orthorhombic (e. g., $Amm2$, $Pna2_{1}$, and $Pmc2_{1}$) and monoclinic (e. g., $C2/m$, $P2/m$, and $P2_{1}/c$) 
crystalline symmetry. All those structures where considered in the four magnetic spin arrangements that can be 
reproduced with our $20$-atoms simulation cell, namely, FM, A-A, A-C, and A-G (see Sec.~\ref{subsec:underP}).  
    
In Fig.~\ref{fig8}, we represent the zero-temperature energy of the crystal phases that according to our DFT 
calculations are energetically most competitive in stoichiometric SCO thin films (some irrelevant magnetic 
order phases have been omitted for clarity) as a function of epitaxial strain, $\eta$ [$\equiv \left(a-a_{0}
\right)/a_{0}$, where $a_{0}$ is equal to $3.89$~\AA]. Interestingly, we found an extreme and complex  
competition between FM and A-A phases, including tetragonal, orthorhombic and monoclinic crystals, at $\eta 
\sim 0$~\% conditions. In particular, the energy of the orthorhombic $Cmmm$ and monoclinic $C2/m$ and $P2_{1}/c$ 
phases, either in the FM or A-A magnetic spin arrangements, are just $\sim 20$~meV/f.u. above that of the 
tetragonal $P4/mbm$ phase (FM case). (We note that none of these low-energy structures was considered 
by Lee and Rabe in works [\onlinecite{lee11,lee11b}].) Such a fierce phase competition appears to be comparable 
to that found in archetypal multiferroic compounds like, for instance, BiFeO$_{3}$~[\onlinecite{dieguez11}]. 
Meanwhile, at small epitaxial strain conditions phases displaying A-C and A-G magnetic spin orders are found 
to be energetically noncompetitive, made the exception of the tetragonal $P4/mbm$ phase (A-C case).      

At $\eta = +2$~\%, we found that the tetragonal ground-state phase undergoes a magnetic phase transition
from FM to A-C (see vertical dashed lines in Fig.~\ref{fig8}). This is a continuous and isostructural  
phase transformation. In order to identify possible electronic-structure changes associated to 
this transition, we calculated the $d$-orbital pDOS in the two relevant $P4/mbm$ phases (see Fig.~\ref{fig9}). 
Our results suggest a change from an intermediate spin state in the FM phase to a low spin state in the 
A-A phase, as in the latter case no energy band gap appears in the spin-down channel. Likewise, along 
the tensile strain-induced phase transition the system is likely to improve its electrical conductivity
properties rather than to becoming an insulator (in contrast to the $P4/mmm~\to~Pmc2_{1}$ transformation 
suggested in~[\onlinecite{lee11,lee11b}], which renders an insulator and ferroelectric compound at $\eta 
\sim +3$~\%). Our predictions appear to be consistent with recent experimental observations reported by 
Callori \emph{et al.}~[\onlinecite{callori15}] and Hu \emph{et al.}~[\onlinecite{hu15}]. Nevertheless, 
new experimental studies characterising in detail the structural and electronic features of SCO thin 
films at moderate tensile strains appear to be necessary for clarifying the discrepancies found between 
ours and previous DFT results~[\onlinecite{lee11,lee11b}].
   
At $\eta = -6.4$~\%, we found that the system transforms into a tetragonal $P4mm$ phase displaying A-C spin 
order and a large $c/a$ ratio of $\sim 1.3$ (see vertical dashed lines in Fig.~\ref{fig8}). In order to 
identify possible electronic-structure changes associated to this transition, we calculated the $d$-orbital 
pDOS in the two involved tetragonal phases (see Fig.~\ref{fig10}). Our analysis reveals that SCO $P4mm$ 
(A-C) thin films are insulator and also ferroelectric, in particular they exhibit a large out-of-plane 
electrical polarisation of $\sim 115$~$\mu$C/m$^{2}$ (as it has been estimated with the methods explained in 
work~[\onlinecite{cazorla15}]). The predicted phase transition, therefore, involves a three-fold structural, 
magnetic, and ferroelectric transformation; however, in spite of its fundamental and technological interests, 
it requires of so large compressive strains that it would be hardly realisable in practice (in contrast to 
the multiferroic phase transformation predicted by Lee and Rabe at $\eta \sim -1$~\% [\onlinecite{lee11,lee11b}]). 
Nonetheless, we found that for the same type of phase transformation to occur in nonstoichiometric SCO thin 
films (i. e., SrCoO$_{2.75}$) a smaller critical epitaxial strain of $\eta = -5.4$~\%~ is needed. Magnetoelectric 
effects then are probably more likely to be observed in nonstoichiometric than in stoichiometric SCO samples.

\section{Summary}
\label{sec:summary}
We have presented a throughout revision of the zero-temperature phase diagram of stoichiometric 
SrCoO$_{3}$ that is obtained with first-principles methods based on density functional theory
(i. e., GGA+U and hybrid functionals). In the bulk case, we have identified a tetragonal $P4/mbm$ 
phase with moderate JT distortions and a $c/a$ ratio of $\sim 1/\sqrt{2}$ as the ground state of 
the system. The same phase remains the most stable as hydrostatic pressures of up to $\sim 70$~GPa 
are applied. This central result either calls for a revision of previous crystallographic analysis 
performed in SCO at low temperatures, or well it represents a failure of DFT methods in describing 
such a highly correlated oxide compound. In SCO thin films, we have found two phase transitions occurring 
at moderate tensile and large compressive epitaxial strains (i. e., $\sim +2$~\% and $\sim -6$~\%, 
respectively). According to our calculations, the first transformation is isostructural and comprises only 
a change in magnetic spin order. The second transition, on the contrary, involves a three-fold structural, 
magnetic, and polar transformation; however, the critical epitaxial strain associated to this transition 
is so large that it appears to be irrealisable in practice. The general description of SCO thin films that 
follows from our study appears to be consistent with recent experimental observations, however, it differs 
considerably from previously reported DFT GGA+U results. The main reason behind such a disagreement is 
likely to be related to the choice of the U value, which as we have shown has an important effect on 
the description of electronic and magnetic spin degrees of freedom. New and systematic experiments 
on stoichiometric SCO, both in bulk and thin film geometries, certainly are necessary in order to advance 
our knowledge of this intriguing material.

\acknowledgments
This research was supported by the Australian Research Council under Future Fellowship
funding scheme (Grant No. FT140100135). Computational resources and technical assistance 
were provided by the Australian Government through Magnus under the National Computational 
Merit Allocation Scheme. P. R. acknowledges an allocation of computing time from the Louisiana
State University High Performance Computing center.

\end{document}